\documentstyle[times,afd-gen,asymetrix,url,ov]{article}

\title{{\LARGE Overview}\EOL
       Computer Supported Query Formulation\\
 {\normalsize Asymetrix Report 94-5}
}
\author{
   H.A. Proper\\
   Asymetrix Research Laboratory\\
   Department of Computer Science\\
   University of Queensland\\
   Australia 4072\\
   E.Proper@acm.org
}
\date{\Version}

   \input{epsf}
   \def\Figurepath{./}
   \def\Scale{0.9}
   \def\epsfsize#./##2{\Scale#./}

\begin{document}
   \maketitle
   {\sc Published as:}
\begin{quote}
  H.A.~(Erik) {Proper}. {An Overview of Computer Supported Query Formulation}. Technical report, Asymetrix Research Laboratory, University of Queensland, Brisbane, Queensland, Australia, 1994.
\end{quote}

   \section{Introduction}
Most present day organisations make use of some automated information system. 
This usually means that a large body of vital corporate information is stored 
in these information systems. 
As a result, an essential function of information systems should be the 
support of disclosure of this information. 

We purposely use the term {\em information disclosure} in this context.
When using the term information disclosure we envision a computer supported
mechanism that allows for an easy and intuitive formulation of queries in a
language that is as close to the user's perception of the universe of discourse
as possible.
From this point of view, it is only obvious that we do not consider a simple 
query mechanism where users have to enter complex queries manually and look 
up what information is stored in a set of relational tables.
Without a set of adequate information disclosure avenues an information system 
becomes worthless since there is no use in storing information that will never 
be retrieved. 

An adequate support for information disclosure, however, is far from a trivial 
problem. 
Most query languages and query mechanisms do not provide any support for the 
users in their quest for information. 
Most of these existing mechanisms can hardly be called disclosure mechanisms 
as they do not provide users any support during the formulation process.
Furthermore, the conceptual schemata of real-life applications tend to be 
quite large and complicated. 
As a result, users may easily become 'lost in conceptual space' and they will 
end up retrieving irrelevant (or even wrong) objects and may miss out on 
relevant objects. 
Retrieving irrelevant objects leads to a low precision, missing relevant 
objects has a negative impact on the {\em recall} 
(\cite{Book:83:Salton:IntroIR}).

The disclosure of information stored in an information system has some clear 
parallels to the disclosure problems encountered in {\em document retrieval 
systems}. 
To draw this parallel in more detail, we quote the information retrieval 
paradigm as introduced in \cite{Report:91:Bruza:StratHypmed}. 
The paradigm starts with an individual or company having an {\em information 
need} they wish to fulfil. 
This need is typically a vague notion and needs to be made more concrete in 
terms of an {\em information request} (the query) in some (formal) language. 
The information request should be as good as possible a description of the 
information need. 
The information request is then passed on to an automated system, or a human 
intermediary, who will then try to fulfil the information request using the 
information stored in the system. 
This is illustrated in the {\em information disclosure}, or {\em information 
retrieval paradigm}, presented in \SRef{\IRParadigm} which is taken from 
\cite{Report:91:Bruza:StratHypmed}.

We now briefly discuss why the information retrieval paradigm for document 
retrieval systems is also applicable for information systems. 
For a more elaborate discussion on the relation between information systems 
and document (information) retrieval systems in the context of the information 
retrieval paradigm, refer to \cite{PhdThesis:94:Proper:EvolvConcModels}. 
In the paradigm, the retrievable information is modelled as a set \Cal{K} of 
{\em information objects} constituting the {\em information base} (or 
population).

In a (multi-media) document retrieval system the information base will be 
a set of multi-media documents (\cite{Book:83:Salton:IntroIR}), while in the 
case of an information system the information base will contain a set of facts 
conforming to a conceptual schema (although this could be multi-media as well). 
Each information object $o \in \Cal{K}$ is {\em characterised} by a set of 
descriptors $\Cal{X}(o)$ that facilitates its disclosure. 
The characterisation of information objects is carried out by a process 
referred to as indexing. 
In an information system, the stored objects (the population or information 
base) can always be identified by a set of (denotable) values, the 
identification of the object. 
For example, an address may be identified as a city name, street name, and 
house number. 
The characterisation of objects in an information system is directly provided 
by the reference schemes of the object types. 

The actual information disclosure is driven by a process referred to as 
{\em matching}. 
In document retrieval applications this matching process tends to be rather 
complex. 
The characterisation of documents is known to be a hard problem 
(\cite{Article:77:Maron:InfRetr}, \cite{Book:86:Craven:SringIndexing}), 
although newly developed approaches turn out to be quite successful 
(\cite{Book:89:Salton:AutomTextProc}). 
In information systems the matching process is less complex as the objects in 
the information base have a more clear characterisation (the identification). 
In this case, the identification of the objects (facts) is simply related to 
the query formulation $q$ by some (formal) query language.
{\def\Scale{0.5} \EpsfFig[\IRParadigm]{The information retrieval paradigm}}

The remaining problem is the query formulation process itself. 
An easy and intuitive way to formulate queries is absolutely essential for an 
adequate information disclosure. 
Quite often, the quest from users to fulfil their information need can be 
aptly described by (\cite{PhdThesis:92:Bruza:IRHypmed}):
\begin{quote} \it 
   I don't know what I'm looking for, but I'll know when I find it.
\end{quote}
In document retrieval systems this problem is attacked by using {\em query by 
navigation} (\cite{Report:91:Bruza:StratHypmed}, 
\cite{PhdThesis:92:Bruza:IRHypmed}) and {\em relevance feedback} mechanisms 
(\cite{Article:89:Rijsbergen:IRLogic}). 
The query by navigation interaction 
mechanism between a searcher and the system is well-known from the Information 
Retrieval field, and has proven to be useful. 
It shall come as no surprise that these mechanisms also apply to the query 
formulation problem for information systems. 
In \cite{Report:92:Burgers:PSMIR}, \cite{Report:93:Burgers:PSMIR}, 
\cite{Report:93:Hofstede:DisclSupport}, 
\cite{PhdThesis:94:Proper:EvolvConcModels} such applications of the 
{\em query by navigation} and {\em relevance feedback} mechanisms have been 
described before. 
When combining the query by navigation and manipulation mechanisms with the 
ideas behind visual interfaces for query formulation as described in e.g. 
\cite{Article:92:Auddino:SUPERVisual} and 
\cite{Article:94:Rosengren:VisualER}, powerfull and intuitive tools for 
computer supported query formulation become feasible, resulting in improved
information disclosure. 
Such tools will also heavily rely on the ideas of direct manipulation 
interfaces (\cite{Article:83:Shneiderman:DirectManip}) as used in present day 
computer interfaces. 

One important step in the improvement of the information disclosure of 
information systems, is the introduction of query languages on a conceptual 
level. 
These languages allow for the formulation of queries in terms common to the
users, i.e. the verbalisations of the types in the conceptual schema.
Examples of such conceptual query languages are RIDL (\cite{Report:82:Meersman:RIDL}), 
LISA-D (\cite{Report:91:Hofstede:LISA-D}, 
\cite{Report:92:Hofstede:LISA-DPromo}), and FORML (\cite{Article:92:Halpin:Subtyping}). 
By letting users formulate queries on a conceptual level, users are 
safeguarded from having to know the exact mapping to internal representations 
(e.g. a set of tables which conform to the relational model) to be able to 
formulate queries in a non conceptual language such as SQL. 
The next step is to introduce ways to support users in the formulation of 
queries in such conceptual query languages (CQL).

   \section{A New Generation of Formulation Mechanisms}
In line with the above discussed information retrieval paradigm and the 
notion of relevance feedback, a query formulation process (both for a 
document retrieval system, and an information system) can be said to roughly 
consist of the following four phases:
\begin{enumerate}
   \item The {\em explorative phase}. 
         What information is there, and what does it mean?
   \item The {\em constructive phase}. 
         Using the results of phase 1, the actual query is formulated.
   \item The {\em feedback phase}. 
         The result from the query formulated in phase 2 may not be completely
         satisfactory. 
         In this case, phases 1 and 2 need to be re-done and the result refined.
   \item The {\em presentation phase}. 
         In most cases, the result of a query needs to be incorporated into a
         report or some other document. 
         This means that the results must be grouped or aggregated in some
         form. 
\end{enumerate}
Depending on the user's knowledge of the system, the importance of the 
respective phases may change. 
For instance, a user who has a good working knowledge of the structure of 
the stored information may not require an elaborate first phase and would 
like to proceed with the second phase as soon as possible. 

In the research for the InfoAssistant product, we try to integrate a palette
of complementary mechanisms to formulate queries on a conceptual level.
These mechanisms are the following:
\begin{description}
   \item[query by navigation]
       This mechanism has been introduced in \cite{Report:92:Burgers:PSMIR}, 
       \cite{Report:93:Proper:DisclSch} and 
       \cite{PhdThesis:94:Proper:EvolvConcModels}.
       The idea behind this mechanism is to shape a conceptual schema, which is
       essentially a graph, as a hypertext and letting users formulate (part
       of) their information need by navigating through this hypertext.
       This mechanisms is particularly suited for those users who do not have
       a clear idea of what information is stored in the information system
       as it is able to truely guide the user through the (structure of the)
       stored information.

       A precursor of the query by navigation mechanism for information 
       systems exists for information retrieval systems 
       (\cite{PhdThesis:92:Bruza:IRHypmed}).
       In expiriments it was shown that in the IR case, this mechanism helps
       novice users in finding their way around the stored information, without
       hampering expert users (\cite{Report:91:Bruza:InfDiscl}).

       All research that remains to be done in this area is some tuning and
       adapting the existing (academic) ideas to the applied situation in
       InfoAssistant.

   \item[query by construction]
       This mechanism has also been discussed before in \cite{Report:92:Burgers:PSMIR}, 
       \cite{Report:93:Proper:DisclSch} and 
       \cite{PhdThesis:94:Proper:EvolvConcModels}.
       This mechanism was born out of the observation that the results of a
       query by navigation sessions are relatively simple queries without
       advanced operations such as grouping, intersections, counting, etc.
       Extending the query by navigation mechanisms with such operations would
       have led to an unacceptable increase in complexity.
       Therefore the introduction of the query by construction as an 
       additional mechanism was chosen.

       The query by construction mechanism is basically a syntax directed 
       editor which allows a user to combine the {\em query particles} resulting
       from {\em query by navigation} (and the three additional mechanisms
       discussed below) sessions to be combined into complex queries using
       the more advanced operations.
    
       Research-wise, this part is finished as
       there is not much research needed for a syntax directed editor

   \item[point to point queries]
       The point to point queries originated from a rough idea from J. Harding.  
       A point to point query starts by selecting two or more object types 
       from a conceptual schema. 
       Then the system should return a list of possible (non cyclic) paths 
       through the information structure between the specified object types. 
       For obvious reasons, the paths in this list should be ordered according 
       to some relevance criterion. 

       This style of querying corresponds to a situation in which users know 
       some aspects (object types) about which they want to be informed, but 
       do not yet know the exact details of their information need and the 
       underlying information structure. 
       The query by navigation mechanism, on the other hand, is intended to 
       support users who do not have an overview of the stored information.

       In \cite{AsyReport:94:Proper:PPQ} this mechanism is discussed and 
       formalised in full detail.

   \item[spider queries]
       This mechanism originated from a discussion with L. Delano. Users
       quite often simply want to know {\em all} information about instances
       of an object type $x$.
       For this purpose the {\em spider queries} were introduced.
       A crucial aspect of spider queries is of course limiting the {\em all
       information} as users probably do not want to be confronted with a
       listing of all information stored in the information system. 

       The idea behind spider queries is to start out from one object type, 
       and to associate all information that is {\em relevant} to this object 
       type. 
       The essential part of a spider query is selecting the object types in 
       the direct suroundings of the initial object type that are considered 
       to be relevant, thus limiting the amount of information returned to the
       user.
 
       This style of querying corresponds to a situation where users only know 
       about the existance of some object types in the conceptual schema about 
       which they would like to be informed. 

       A complete discussion and formal treatment of this mechanism can be
       found in \cite{AsyReport:94:Proper:PPQ}.

   \item[natural language queries]
       A more commonly known mechanism for computer supported query formulation
       are (semi) natural language query formulation systems.
       These mechanisms try to interpret sentences in a semi-natural language
       format and generate an appropriate query in SQL.
       
       Our aim is to try and integrate these ideas with the newly added
       formulation mechanisms.
       One important aspect of this integration is that it would allow us to
       interpret the natural language sentence, and then automatically formulate
       a query in a conceptual query language rather than SQL.
       This would certainly put the user in a much better position to 
       validate the resulting query than to confront users with an SQL query.

       A natural language formulation mechanism is usefull for those users 
       who know what information is stored in the information system, but who
       do not know the exact names of the types.
       The flexibility of a semi-natural language would then cater for this.
\end{description}
In the remainder of this overview report we discuss some example session
using the different disclosure avenues.
This should give a more hands-on idea of what these mechanisms are about.

   \section{An Example Session}

In this section we discuss a sample session using the query formulation
component of InfoAssisant.
The discussed example operates on a conceptual schema for the administration 
of the election of American presidents. 
The example schema itself is not shown; the structure of the domain will 
become clear from the sample session. 
Note that the quality of the verbalisations of paths expressions used in the 
examples in this section should be improved.
However, this is the subject of further research. 

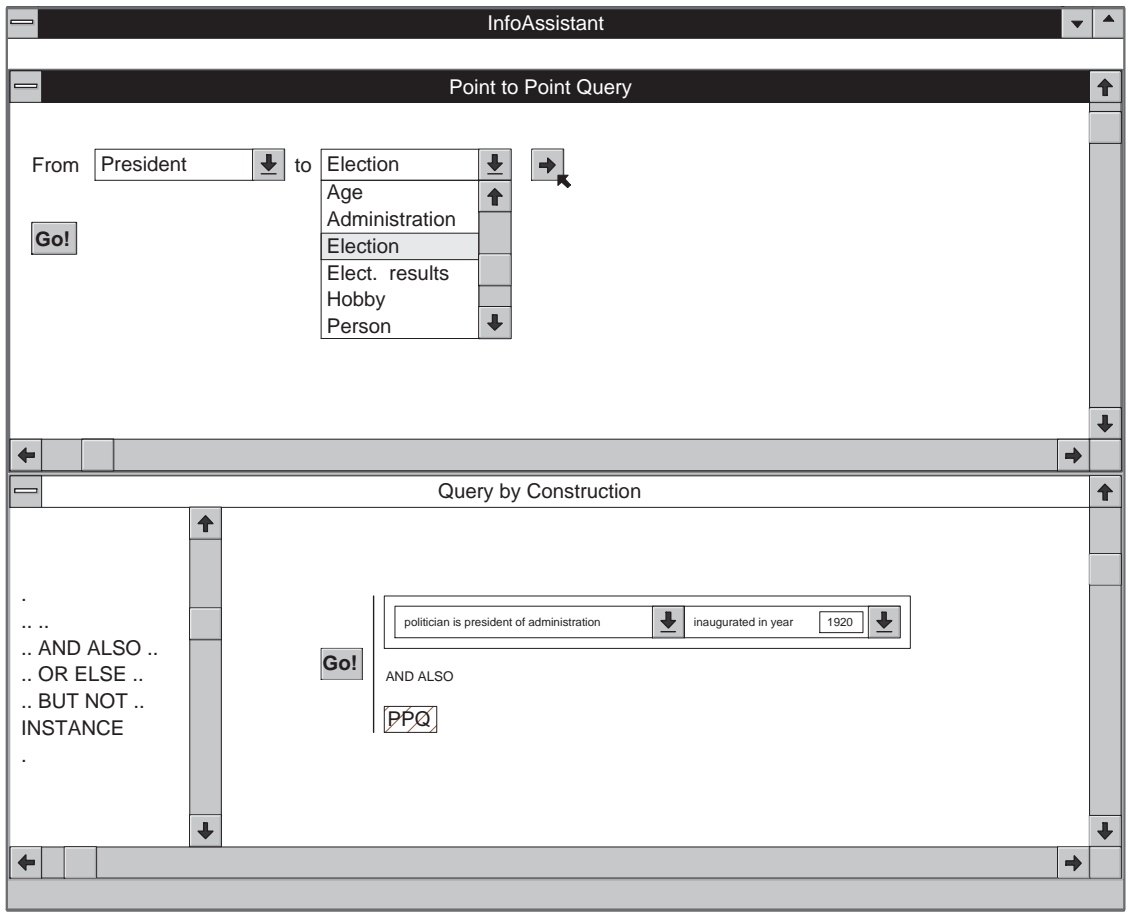
{\def\Scale{0.70} \EpsfFig[\PPQStart]{Building a PPQ query}}
In \SRef{\PPQStart}, a possible screen is depicted for building queries using 
a point to point query mechanism. 
The upper window is concerned with the point to point query itself, whereas 
the lower window contains the complete query under construction. 
When specifying a point to point query a user specifies a sequence of object 
types: the points. 
For each point, the user is offered a listbox containing all object types 
present in the conceptual schema. 
The order of the object types in the listbox should 
preferably be based on some notion of conceptual importance 
(\cite{Article:94:Campbell:Abstraction}). 
In \SRef{\PPQExtension} an existing point to point query path from president 
to election is extended with another point.
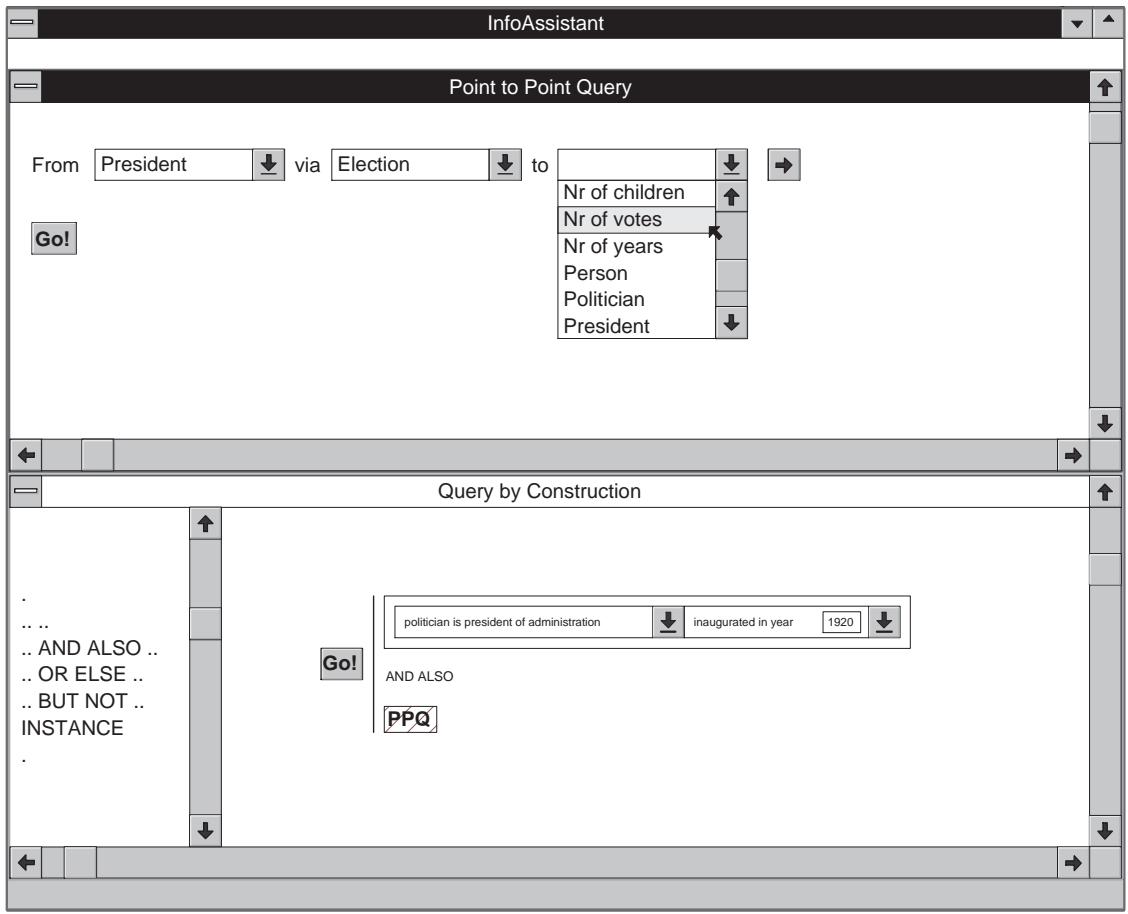
{\def\Scale{0.70} \EpsfFig[\PPQExtension]{Extending the PPQ path}} 

{\def\Scale{0.70} \EpsfFig[\PPQComplete]{Completing a PPQ}}
After all points of the point to point query have been specified, the point 
to point query can be transformed into a proper query (i.e. a path through 
the conceptual schema) by pressing the \SF{Go!} button in the point to point 
query window. 
In \SRef{\PPQComplete}, this process is illustrated. 
The sample PPQ involves three points. 
Therefore, two paths through the conceptual schema will result. 
We now shift our attention from the point to point query window to the query 
by construction window. 
Note that the small box containing the \SF{PPQ} abbreviation is now replaced 
by the paths resulting from the point to point query (i.e. \SF{President 
winning election which resulted in nr of votes}). 
The system initially inserts a most likely path. 
The user can, however, select alternative paths using a listbox. 
Note that not all alternative paths between the two points are listed in 
the listbox. 
The reason for this is the NP completeness of the path searching problem. 
To avoid the NP completeness problem, only the best paths are listed 
initially. 
However, potentially all paths can be selected (which still remains NP 
complete) by repeatedly selecting the \SF{MORE} option. 
In the remainder of this article we will discuss this in more detail.

{\def\Scale{0.70} \EpsfFig[\PPQNavigate]{Switching to query by navigation}}
Since every path resulting from a query by navigation session connects two 
points in the conceptual schema, any path through the conceptual schema 
displayed in the query by construction screen can be used as a starting point
for a query by navigation session, and vice versa. 
This is illustrated in \SRef{\PPQNavigate}. 
In this session, the user has selected the box which contains the two paths 
\SF{politician is president of administration} and \SF{inaugurated in year} 
for a query by navigation session. 
The upper window now displays a node in the query by navigation session, with 
the path \SF{politician is president of administration inaugurated in year}
as its focus. 
If the user had selected the \SF{inaugurated in year} listbox, the 
initial focus would have been \SF{administration inaugurated in year}.

The query by construction window in \SRef{\PPQNavigate} basically offers a 
syntax directed editor. 
In the left part of the window all possible constructs from the query 
language are listed. 
In our examples we have used the constructs defined in LISA-D. 
Once the FORML and LISA-D languages have been merged, a more complete 
language for the query by construction part will result.

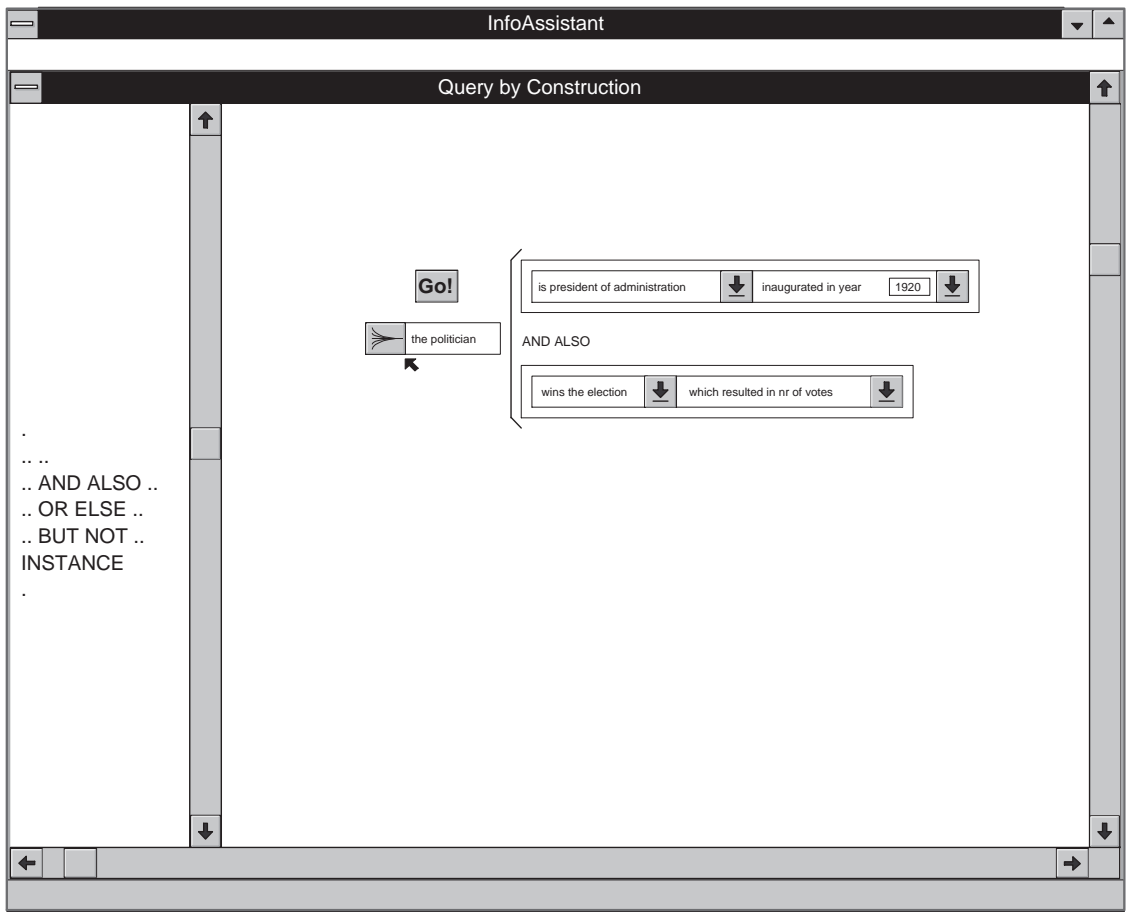
{\def\Scale{0.70} \EpsfFig[\SQStart]{Start of a spider query}}
Next we discuss a session involving a spider query.
We start out from an existing query in a query by construction window, which
could have been constructed using a query by navigation querie or a point to
point query. 
Note that this could also be single object type, e.g. politician. 
The spider query mechanism adds one important aspect to the query 
by construction window, the spider button: \SpiderButton. 
When a user presses this button, the system calculates the spider query of 
the object type directly to the right of the button. 
This is illustrated in \SRef{\SQResult}. 
The system allows for the removal of parts of the resulting spider query that 
are not considered to be relevant by the user.
Suppose the user is not interested in \SF{administration is headed by} and 
\SF{election won by}, then these paths can be deleted, which leads to the 
screen depicted in \SRef{\SQResultPruned}.
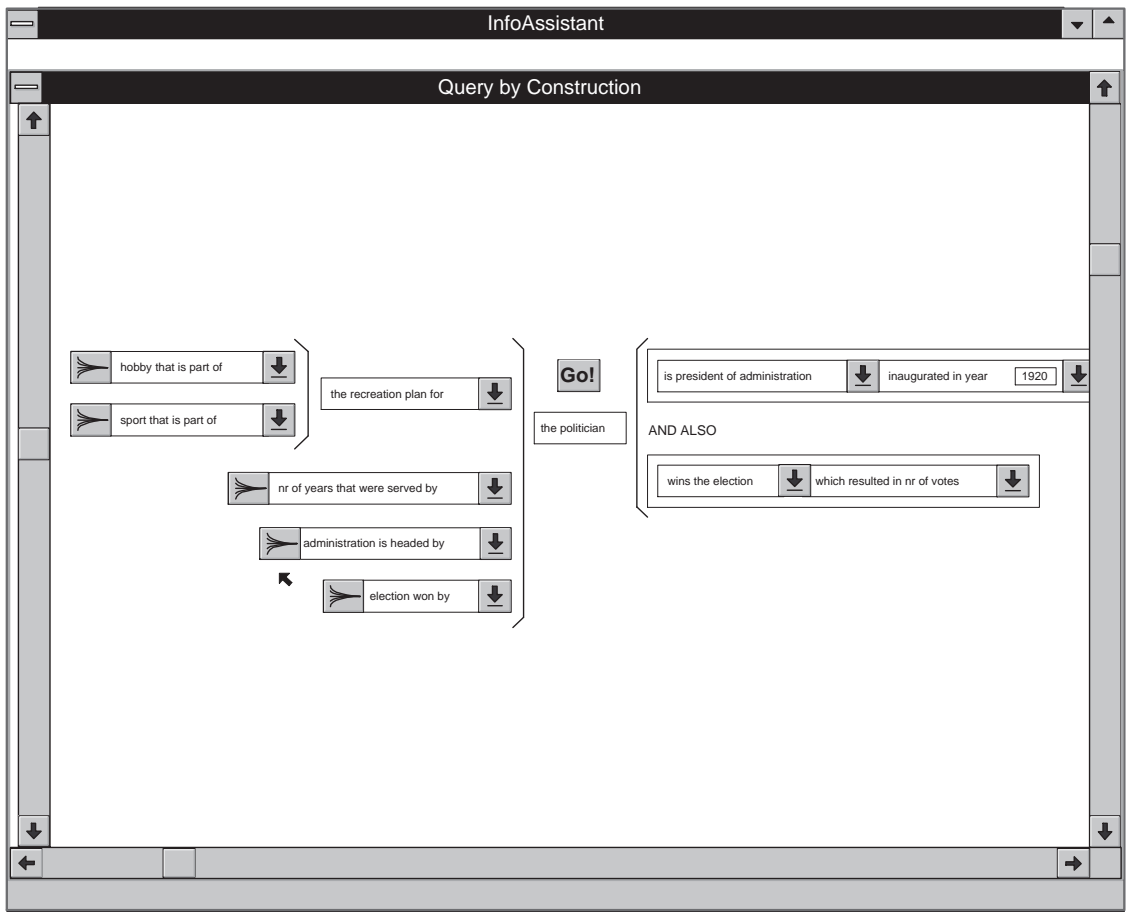
{\def\Scale{0.70} \EpsfFig[\SQResult]{Result of a Spider Query}}

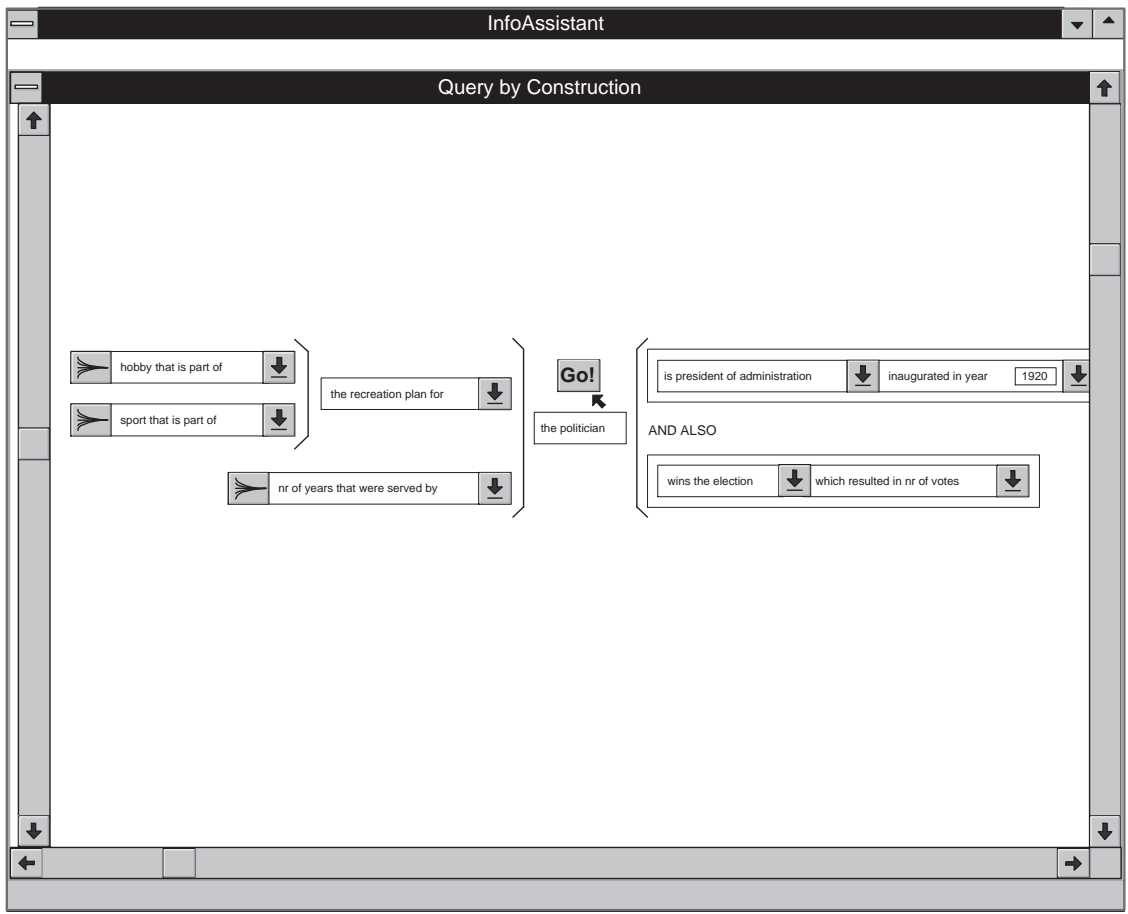
{\def\Scale{0.70} \EpsfFig[\SQResultPruned]{Pruning the Spider Query}}
It is now interesting to see that a query essentially is a double tree with a shared 
root (politician in the example).
Furthermore, the leaves on the tree resulting from the spider query can be extended 
further if desired by commencing new spider queries.
Finally, since the result of a spider query is constructed from path expressions 
as well, these expressions have the \PPQButton~associated that can be used to 
select alternative paths between the head and tail object types. 
Furthermore, the paths can also be used as a starting point of a query by 
navigation session. 
This latter posibility is illustrated in \SRef{\SQQBN}.
{\def\Scale{0.70} \EpsfFig[\SQQBN]{Switching to query by navigation}}

   \AddBib{asy}
   \BIBLIOGRAPHY{alpha}
\end{document}